\documentclass[11pt]{paper}

\usepackage{amsmath}

\usepackage{times}
\usepackage{bm}
\usepackage{natbib}

\usepackage[plain,noend]{algorithm2e}

\usepackage[top=1in, bottom=1.5in, left=1.3in, right=1.3in]{geometry} 

\def\~ka{{\it Biometrika}}

\def\var{\textup{var}}

\def\corr{\textup{corr}}
\def\diff{\textup{d}}
\def\R{\textup{R}}
\def\dN{\textup{N}}
\def\pr{\textup{pr}}

\def\BCV{\textsc{bcv}}
\def\MISE{\textsc{mise}}
\def\AMISE{\textsc{amise}}
\def\mAMISE{{\small \textup{m}}\textsc{amise}}
\def\mBCV{{\small \textup{m}}\textsc{bcv}}
\def\IQR{\textsc{iqr}}
\def\SJ{\textsc{sj}}
\def\mSJ{{\small \textup{m}}\textsc{sj}}
\def\ISE{\textsc{ise}}

\newtheorem{theorem}{Theorem}

\newtheorem{corollary}[theorem]{Corollary}
\newtheorem{lemma}[theorem]{Lemma}	
\newtheorem{condition}[theorem]{Condition}
\newtheorem{definition}[theorem]{Definition}

\usepackage{amssymb}
\usepackage{multirow}

\begin{document}

%
%

\title{Bandwidth Selection for Kernel Density Estimation with a Markov Chain Monte Carlo Sample}

\author{\textup{\large Hang J. Kim} \\
{\small Department of Mathematical Sciences, University of Cincinnati, Cincinnati, OH 45221, U.S.A.} \\
\textup{\small hang.kim@uc.edu} \\
\textup{\large Steven N. MacEachern} \\
{\small Department of Statistics, The Ohio State University, Columbus, OH 43210, U.S.A.} \\
\textup{\small snm@stat.osu.edu} \\
\textup{\large Yoonsuh Jung} \\
{\small Department of Mathematics and Statistics, University of Waikato, Hamilton 3240, New Zealand} \\\textup{\small yoonsuh@waikato.ac.nz}
}

%
%

\maketitle

\begin{abstract}
Markov chain Monte Carlo samplers produce dependent streams of variates drawn from the limiting distribution of the Markov chain.  With this as motivation, we introduce novel univariate kernel density estimators which are appropriate for the stationary sequences of dependent variates.  We modify the asymptotic mean integrated squared error criterion to account for dependence and find that the modified criterion suggests data-driven adjustments to standard bandwidth selection methods.  Simulation studies show that our proposed methods find bandwidths close to the optimal value while standard methods lead to smaller bandwidths and hence to undersmoothed density estimates.  Empirically, the proposed methods have considerably smaller integrated mean squared error than do standard methods.

\end{abstract}


\section{Introduction}
Kernel density estimation has been extensively studied since the early works of \cite{Parzen:1962} and \cite{Rosenblatt:1971}.  Theoretical developments \citep{Rudemo:1982,Bowman:1984,Silverman:1986,Scott:1987,Park:1990,Hall:1991,Sheather:1991} have been coupled with practical guidance on implementation of the methods \citep{Jones:1995,Sheather:2004}, and these density estimates are now used wherever data is collected--from archeology, to economics, to genetics, and beyond. 
Particular attention has been given to selection of the kernel's bandwidth, and the development of automatic methods of bandwidth selection has put kernel density estimation in all of the major statistical software packages, for example, in PROC KDE in SAS, and in the `stats' package in R.  The most commonly used methods for the choice of the bandwidth involve cross-validation or rely on plug-in approaches.

The bulk of the literature on kernel density estimation assumes that the observations have arisen as an independent sample from some unknown distribution, but there have been a modest number of studies on the asymptotic properties of the kernel density estimator when the data are dependent. For example, \cite{Yakowitz:1989} showed that the kernel density estimator at an evaluation point is asymptotically normal when the sample is from a stationary time series. Regarding the data-driven bandwidth selection approach, \cite{Hart:1990} found that the ordinary cross-validation method is asymptotically optimal for weakly dependent data in terms of rate, as in the independent case. \cite{Hall:1995} further studied asymptotic properties of the optimal bandwidth under different levels of dependence. 
However, despite the vast literature on asymptotic properties of the estimator, there are few studies to provide practical guidance for bandwidth selection with data generated from a dependent process. 

In this paper, we propose data-driven bandwidth selection methods in one-dimensional kernel density estimation when the data are dependent.  Although our approach can be applied to any dependent data which satisfy certain mixing conditions, we specifically focus on kernel density estimation for samples generated from a Markov chain Monte Carlo algorithm. 

Markov chain Monte Carlo algorithms have been at the core of modern Bayesian analysis since the seminal
work of \cite{Geman:1984} and \cite{Gelfand:1990}.  They are used to produce a dependent sequence of variates from the posterior distribution.  These variates are then used to make a formal inference and to summarize the posterior distribution informally. 
The summary of the posterior distribution is typically accomplished via a kernel density estimate. Despite the importance and popularity of kernel smoothing to summarize posterior distributions 
\citep[e.g.,][]{Hoti:2002,Yi:2003,Mathew:2012}, we have been unable to find any studies on kernel bandwidth selection with a Markov chain Monte Carlo sample. Instead, due to the lack of practical bandwidth selection rules, the bandwidth is subjectively chosen by the analyst.

The main idea of this paper is to rewrite the asymptotic mean integrated squared error so that its leading terms include a measure of dependence referred to as the integrated autocorrelation time. Based on the modified asymptotic mean integrated squared error, we suggest modified versions of biased cross-validation and two Sheather-Jones plug-in methods. In simulation studies where the data are drawn from a Markov chain Monte Carlo algorithm, we show that the proposed methods find bandwidths close to the optimal value, while the standard methods result in undersmoothed estimates.

The remainder of the paper is organized as follows. Section \ref{sec:KED} reviews the basic theorems of kernel density estimation under an independence assumption and introduces two popular bandwidth selection methods. Section \ref{sec:MCMC} presents theoretical results for kernel density estimation under Markov chain Monte Carlo samples. Section \ref{sec:suggested} suggests bandwidth selection methods for the dependent sample, followed by Section \ref{sec:simulation} where simulation studies compare the proposed methods to their original versions. Section \ref{sec:discussion} concludes with a brief discussion. 

\section{Background} \label{sec:KED}

\subsection{Basic setting for kernel density estimation}

Let $\{Y_1,\ldots,Y_n\}$ be a sample from an unknown density $f$.
The kernel density estimator of $f$ at the evaluation point $x$ is defined as
\begin{equation*} \label{eq:KDE}
	\hat{f}_h(x) = \frac{1}{n h} \sum_{i=1}^n K \left( \frac{x-Y_i}{h} \right), 
\end{equation*}
where the kernel $K$ is generally chosen to be a symmetric probability density and $h$ is a smoothing parameter, referred to as the bandwidth. The performance of the kernel density estimator mainly depends on the selection of a bandwidth within a class of kernels rather than on the kernel's shape \citep{Sheather:2004,Scott:2015}. 

Data-driven bandwidth selection is often motivated by the desire to minimize the mean integrated squared error of the estimator
\begin{align} 
 \MISE(h) & = \int E \left\{ \hat{f}_h(x)-f(x) \right\}^2 \diff x 	= \int \var \left\{ \hat{f}_h(x) \right\} \diff x + \int \left[ E\{\hat{f}_h(x)\}-f(x) \right]^2 \diff x. \label{eq:MISE}
\end{align}
The second term of the mean integrated squared error in \eqref{eq:MISE} is the integrated squared bias.
Under the following conditions commonly assumed in the kernel density estimation literature \citep[e.g.,][]{Scott:1985,Silverman:1986,Scott:1987}:
\begin{condition} \label{cond:C1}
$|t|^{r+1} K(t) \rightarrow 0$ as $|t| \rightarrow \infty$,
\end{condition}
\begin{condition} \label{cond:C2}
$\int |t|^r K(t) dt < \infty$,
\end{condition}
\begin{condition} \label{cond:C3}
$f^{(r)} \in L^1$, i.e., $ \int |f^{(r)}(x)| \diff x  < \infty$,
\end{condition}
\begin{condition} \label{cond:C4}
$f^{(r)}$ is continuous,
\end{condition}
\begin{condition} \label{cond:C5}
$K$ is a symmetric probability density with mean 0 and finite variance,
\end{condition}
the integrated squared bias term can be approximated up to order $h^4$ irrespective of the assumptions of dependence or independence of the sample.
\begin{theorem} \label{thm:int_bias_sq} 
If Conditions \ref{cond:C1}--\ref{cond:C5} hold for $r=2$, then
\begin{equation*} \label{eq:MISE_bias2}
\int \left[ E\{\hat{f}_h(x)\}-f(x) \right]^2 \diff x = \frac{h^4}{4} \mu_2^2 \
 \R(f^{''}) + o(h^4)
\end{equation*}
as $h \rightarrow 0$, where $\R$ denotes the squared $L_2$ norm of a function, i.e., $\R(\nu) = \int \nu^2(u) \diff u$ and $\mu_r$ denotes the $r$th moment of the kernel, i.e., $\mu_r = \int u^r K(u) \diff u$. 
\end{theorem}

\subsection{Bandwidth selection methods under independence} \label{sec:independence}

The first term of the mean integrated squared error in \eqref{eq:MISE} is the integrated variance.  Under the independence assumption, the integrated variance term can be approximated up to order $n^{-1}$. 
\begin{theorem} \label{thm:int_var_indep}
Suppose that Conditions \ref{cond:C1}--\ref{cond:C5} hold for $r=1, 2$. If $Y_1,\ldots,Y_n$ are independent random variables from $f$, then 
\begin{equation*} \label{eq:MISE_var}
\int \var \left\{ \hat{f}_h(x) \right\} \diff x = \frac{1}{nh} \R(K) + O \left( n^{-1} \right)
\end{equation*}
as $n \rightarrow \infty$. 
\end{theorem}
When $\{Y_1,\ldots,Y_n\}$ is an independent sample, the asymptotic mean integrated squared error is defined by the two leading terms,  
\begin{equation} \label{def:AMISE}
\AMISE(h) = \frac{1}{nh} \R(K) + \frac{h^4}{4} \mu_2^2 \R(f^{''}) ,
\end{equation}
since 
 $ \MISE(h) = \AMISE(h) + O \left(n^{-1} + h^5 \right) $
 as $n \rightarrow \infty$, $h=h(n) \rightarrow 0$, and $n h(n) \rightarrow \infty$, by Theorems \ref{thm:int_bias_sq} and \ref{thm:int_var_indep}. 
Then, the optimal bandwidth with regard to the asymptotic mean integrated squared error is calculated by setting its first derivative equal to zero,
\begin{equation*} \label{eq:hAMISE}
  h = \left\{ \frac{1}{n} \frac{\R(K)}{\mu_2^2 \ \R(f^{''})} \right\}^{\frac{1}{5}}.
\end{equation*}

This paper focuses on some popular bandwidth selection approaches: the biased cross-validation method and two types of the Sheather-Jones plug-in methods. 
\cite{Scott:1987} proposed the biased cross-validation method where the bandwidth is chosen to minimize an objective function instead of the asymptotic mean integrated squared error. Specifically, the unknown quantity $\R(f^{''})$ in \eqref{def:AMISE} is replaced with its estimate, leading to the biased cross-validation objective function 
\begin{equation} \label{eq:BCV_objective}
\BCV(h) = \frac{1}{nh} \R(K) + \frac{h^4}{4} \mu_2^2  \left\{ \R(\hat{f}^{''}_h)- \frac{\R(K^{''})}{nh^5} \right\},  
\end{equation} 
where $\hat{f}^{''}_h$ is the second derivative of the kernel density estimate $\hat{f}_h$.  

\cite{Sheather:1991} suggested several plug-in methods \citep{Park:1990, Jones:1991} that use another bandwidth $g$, different from  $h$, to estimate $\R(f^{''})$. The first approach, referred to as the solve-the-equation method, finds the solution of
\begin{equation} \label{eq:h_hat_2S}
  h = \left\{ \frac{1}{n} \frac{\R(K)}{\mu_2^2 S\{g(h)\}} \right\}^{\frac{1}{5}}.
\end{equation}
Here, $S\{g(h)\}$ is the estimator of $\R(f^{''})$ based on rules of thumb devised for the normal distribution.  The function is 
\begin{equation*} \label{eq:S_D}
S\{g(h)\} = \frac{1}{n(n-1) \ g(h)^5}  \ \sum_{i=1}^n \sum_{j=1}^n \phi^{(4)}\left\{ \frac{Y_i-Y_j}{g(h)} \right\},
\end{equation*}
where $\phi$ denotes the standard normal density such that $\phi(u) =(2\pi)^{-1/2} \exp(-u^2/2)$ and $\phi^{(r)}$ denotes its $r$th derivative. \cite{Sheather:1991} solved \eqref{eq:h_hat_2S} after replacing $g(h)$ with
\begin{equation} \label{eq:hat_g_h}
\hat{g}(h)  = \text{1$\cdot$357} \left\{ \frac{S(a)}{T(b)} \right\}^{\frac{1}{7}} h^{\frac{5}{7}},
\end{equation}
where
$$
S(a) = \frac{1}{n(n-1) \ a^5} \sum_{i=1}^n \sum_{j=1}^n \phi^{(4)} \left( \frac{Y_i-Y_j}{a} \right), \   T(b) = - \frac{1}{n(n-1) \ b^7} \sum_{i=1}^n \sum_{j=1}^n \phi^{(6)} \left( \frac{Y_i-Y_j}{b} \right).
$$
The smoothing parameters are computed as $a=$ 0$\cdot$920 $n^{-1/7} \IQR$ and $b=$ 0$\cdot$912 $ n^{-1/9} \IQR $ where the symbol $\IQR$ denotes the sample interquartile range of $\{Y_1,\ldots,Y_n\}$. 

For comparison, we also introduce another approach of \cite{Sheather:1991} that finds $h$ by minimizing the objective function
\begin{equation} \label{eq:h_hat_2M}
	\SJ(h) = \frac{1}{nh} \R(K) + \frac{h^4}{4} \mu_2^2 S\{\hat{g}(h)\}.
\end{equation}

\section{Kernel density estimation with a Markov chain Monte Carlo sample} \label{sec:MCMC}

\subsection{Mixing conditions and integrated autocorrelation time}

The usefulness of a Markov chain Monte Carlo method is enhanced by asymptotic unbiasedness and a fast rate of convergence. Let $\{Y_1,\ldots,Y_n\}$ be a Harris ergodic Markov chain with an invariant distribution $f$. When $E|f| < \infty$, the ergodic theorem guarantees asymptotic unbiasedness, i.e., $\sum_{i=1}^n \nu(Y_i) / n \rightarrow E (\nu) \text{ as } n \rightarrow \infty$ with probability one, for any initial distribution. The convergence rate is closely connected to various mixing conditions of a Markov chain.

The integrated autocorrelation time is a measure of dependence defined by
\begin{equation} \label{eq:IAT}
 \tau_n = \sum_{t=-(n-1)}^{n-1} \left(1-\frac{|t|}{n} \right) \corr (Y_0,Y_t).
\end{equation}
In the context of kernel density estimation, we define the integrated autocorrelation time of the kernel as
\begin{equation} \label{eq:IAT_K}
	\tau_n(K_{h,x}) = \sum_{t=-(n-1)}^{n-1} \left(1-\frac{|t|}{n} \right) \corr \left\{K_h(x-Y_1),K_h(x-Y_{t+1})\right\}, 
\end{equation}
where $K_h(u) = K(u/h)/h$. 
The variance of the kernel density estimator can be expressed in terms of the integrated autocorrelation time, 
\begin{equation*} \label{eq:newvar}
\var \left\{ \hat{f}_h(x) \right\} = \frac{1}{n^2} \var \left\{ \sum_{i=1}^n K_h(x-Y_i) \right\} 
= \frac{1}{n} \var \left\{ K_h(x-Y_1) \right\} \tau_n(K_{h,x}).
\end{equation*}
Unlike the standard version of the integrated autocorrelation time in \eqref{eq:IAT}, the integrated autocorrelation time of the kernel in \eqref{eq:IAT_K} applies to a sequence of functions, $K_h$,  which change with $h$ and so with $n$.  Thus the asymptotic properties need to be investigated with limiting values of $h$ and $n$. 

The following theorem shows that for a Harris ergodic Markov chain the integrated autocorrelation time of the kernel
increases as $n$ increases.
\begin{theorem} \label{thm:tau_K}
Suppose that Conditions \ref{cond:C1}--\ref{cond:C5} hold for $r=1, 2$. If $\{Y_1, \ldots, Y_n\}$ is a Harris ergodic Markov chain with an invariant distribution $f(\cdot)$ on a state space $\mathcal{Y}$, then for some $\delta >0$,  
$\tau_n(K_{h,x}) = O \left( h^{-\delta/(2+\delta)} \right)$ 
almost everywhere with regard to $x$ as $n \rightarrow \infty$.
\end{theorem}
If the Markov chain has a faster convergence rate, for example, it is geometrically ergodic,
we can show that the integrated autocorrelation time of the kernel is finite as follows. 
\begin{theorem} \label{thm:IACT_geo}
If $\{Y_1,\ldots,Y_n\}$ is a geometrically ergodic chain and its Markov transition kernel $P$ satisfies detailed balance with respect to $f$, i.e., 
$$
	f(\diff y) P(y, \diff y') = f( \diff y') P(y', \diff y),  \quad y, y' \in \mathcal{Y}, 
$$
then 
$ \tau_n(K_{h,x})=O(1)$ almost everywhere with regard to $x$ as $n \rightarrow \infty$.
\end{theorem}
Theorem \ref{thm:IACT_geo} is useful when dealing with a sample from a standard Markov chain Monte Carlo method, such as a Metroplis-Hastings chain or a Gibbs sampler which are guaranteed to produce geometrically
ergodic chains \citep{Chan:1993,Roberts:1996,Jones:2004}.
  
\subsection{Density estimation with a Markov chain}

In this section, we suggest a modified version of mean integrated squared error that is appropriate for a sample of dependent data and show its asymptotic properties.  For a Harris ergodic Markov chain, the mean integrated squared error in \eqref{eq:MISE} is approximated with the integrated autocorrelation time of the kernel which reflects the dependence in the sample as in the following theorem.
\begin{theorem} \label{thm:MISE_MCMC}
Suppose that Conditions \ref{cond:C1}--\ref{cond:C5} hold.  If $\{Y_1,\ldots,Y_n\}$ is a Harris ergodic Markov chain with an invariant distribution $f$ on a state space $\mathcal{Y}$, then 
\begin{equation*} \label{eq:MISE_var_dep}
\int \var \left\{ \hat{f}_h(x) \right\} \diff x = \frac{1}{nh} \R(K) \zeta_f(K_h) + O \left( n^{-1} h^{-\frac{\delta}{2+\delta}} \right)
\end{equation*}
as $n \rightarrow \infty$ for some $\delta > 0$, where $\zeta_f(K_h) = \int \tau_n(K_{h,x}) f(x) \diff x$. Therefore, combined with Theorem \ref{thm:int_bias_sq},
\begin{equation} \label{eq:MISE_dep}
 \MISE(h) = \frac{1}{nh} \R(K) \zeta_f(K_h) + \frac{h^4}{4} \mu_2^2 \
 \R(f^{''}) + O \left(n^{-1} h^{-\frac{\delta}{2+\delta}} + h^5 \right)
\end{equation}
as $n \rightarrow \infty$.
\end{theorem}
Retaining only the leading terms, we define the modified version of the asymptotic mean integrated squared error.
\begin{definition}
\begin{equation} \label{def:modifiedAMISE}
\mAMISE(h) = \frac{1}{nh} \R(K) \zeta_f(K_h) + \frac{h^4}{4} \mu_2^2 \R(f^{''}).
\end{equation}
\end{definition}
In the modified form, $\zeta_f(K_h)$ is multiplied by the first term of the original asymptotic mean integrated squared error in \eqref{def:AMISE}, reflecting the dependence in the sample. When $Y_1,\ldots,Y_n$ are independent random variables, the modified asymptotic mean integrated squared error in \eqref{def:modifiedAMISE} is the same as in \eqref{def:AMISE} since independence implies $\tau_n(K_{h,x})=1$ followed by $\zeta_f(K_h)=1$.

\begin{corollary} \label{cor:MISE_mAMISE:dep}
Suppose that Conditions \ref{cond:C1}--\ref{cond:C5} hold.  If $\{Y_1,\ldots,Y_n\}$ is a Harris ergodic Markov chain with an invariant distribution $f$ on a state space $\mathcal{Y}$, then
$$
\MISE(h) = \mAMISE(h) + O \left(n^{-1} h^{-\frac{\delta}{2+\delta}} + h^5 \right)
$$
as $n \rightarrow \infty$.
\end{corollary}
\begin{corollary} \label{cor:geo}
If $\{Y_1,\ldots,Y_n\}$ is a geometrically ergodic chain and the Markov transition kernel $P$ satisfies detailed balance with respect to $f$, then the equation \eqref{eq:MISE_dep} holds with $\delta=0$, i.e., 
$$
\MISE(h) = \mAMISE(h) + O \left(n^{-1} + h^5 \right)
$$
as $n \rightarrow \infty$.
\end{corollary}

\section{Suggested bandwidth selection methods} \label{sec:suggested}

\subsection{Modified biased cross-validation method}
The biased cross-validation method for bandwidth selection can be modified by replacing the objective function \eqref{eq:BCV_objective} with one which accounts for dependence in the sample.  For this approach, we suggest selecting the bandwidth which minimizes
\begin{equation} \label{eq:mBCV}
\mBCV(h) = \frac{1}{nh} \R(K) \zeta_{\hat{f}_h}(K_h) + 
\frac{h^4}{4} \mu_2^2 \left\{ \R(\hat{f}^{''}_h)- \frac{\R(K^{''})}{nh^5} \right\},
\end{equation} 
where $\zeta_{\hat{f}_h}(K_h)=\int \tau_n(K_{h,x}) \hat{f}_h(x) \diff x$ is introduced to reflect the influence of dependence on the variance.
The second term in \eqref{eq:mBCV} is the same with that in \eqref{eq:BCV_objective}, which is an estimate for the leading term of the integrated squared bias in Theorem \ref{thm:int_bias_sq}. For the asymptotic properties of the second term, we refer readers to \cite{Scott:1987}.
If the Markov chain is geometrically ergodic, the modified objective function in \eqref{eq:mBCV} converges to the modified asymptotic mean integrated squared error in \eqref{def:modifiedAMISE} by the following theorem.
\begin{theorem} \label{thm:conv_zeta_f}
Suppose $\{Y_1,\ldots,Y_n\}$ is geometrically ergodic. If $nh (\log n)^{-3} \rightarrow \infty$, i.e., $h$ tends to zero sufficiently  slowly, then $\zeta_{\hat{f}_h}(K_h)$ converges to $\zeta_f(K_h)$ almost surely as $n \rightarrow \infty$. 
\end{theorem}

With observed data, we compute the modified objective function by plugging-in the sample integrated autocorrelation time of the kernel in the place of $\tau_n(K_{h,x})$ in \eqref{eq:mBCV}, which is defined by 
\begin{equation*} \label{eq:sampleIAT_K}
	\hat{\tau}_n(K_{h,x}) = \sum_{t=-(n-1)}^{n-1} \left(1-\frac{|t|}{n} \right) \rho \left\{K_h(x-Y_1),K_h(x-Y_{t+1})\right\},
\end{equation*}
where $\rho(Y_1,Y_{t+1})$ denotes the sample autocorrelation at lag $t$ in $\{Y_1,\ldots,Y_n\}$.

\subsection{Modified Sheather-Jones plug-in methods}

We suggest two modified versions of the Sheather-Jones plug-in methods. The solve-the-equation method  modifies \eqref{eq:h_hat_2S} and solves 
\begin{equation*} \label{eq:h_mSJmin}
  h = \left\{ \frac{1}{n} \frac{\R(K) \zeta_{\hat{f}_h}(K_h)}{\mu_2^2 S\{\hat{g}(h)\}} \right\}^{\frac{1}{5}} 
\end{equation*}
for $h$.
The second replaces \eqref{eq:h_hat_2M} with
\begin{equation} \label{eq:h_mSJmin_obj}
	\mSJ(h) = \frac{1}{nh} \R(K) \zeta_{\hat{f}_h}(K_h) + \frac{h^4}{4} \mu_2^2 S\{\hat{g}(h)\},
\end{equation}
where $\hat{g}(h)$ is the estimate of $g(h)$ as defined in \eqref{eq:hat_g_h}. The selected bandwidth is the minimizer of \eqref{eq:h_mSJmin_obj}.  Theorem \ref{thm:conv_zeta_f} provides the asymptotic properties of the first term in \eqref{eq:h_mSJmin_obj}.  The second term is identical to that in \eqref{eq:h_hat_2M},  and \cite{Sheather:1991} describe its asymptotic behavior.

\section{Simulation study} \label{sec:simulation}

The simulation study in this section illustrates the performance of the proposed methods under independent samples and under Markov chain samples.  Two sampling methods are considered.  The first is independent draws from a distribution, where we expect traditional kernel density estimates to perform well.  The second is Markov chain Monte Carlo draws from a Metropolis-Hastings chain, where we expect the proposed methods to do well.  The Metropolis-Hastings chains are initialized in the limiting distributions.  Proposals are random walk proposals with Gaussian increments.  The standard deviation of the increment is chosen so that the acceptance rate is between 0$\cdot$2 and 0$\cdot$25.  This led to sample integrated autocorrelation times ranging from 5$\cdot$3 to 10$\cdot$7.  Three distributions are included in the study:  a normal distribution $\dN(3,2^2)$ with mean $3$ and variance $4$, a two-component mixture of normals $0.7 \dN(0,1^2)+0.3\dN(4,1^2)$, and a log-normal distribution whose mean is $\exp(1+$ 0$\cdot$3$^2/2)$, with a corresponding normal distribution having mean $1$ and variance 0$\cdot$09. These are examples of symmetric, multimodal, and skewed distributions, respectively.  For each sampling method-distribution combination, a sample of size $10,000$ was drawn.  This process was repeated $50$ times.

For each sample in each replicate, we compared several kernel density estimation methods.  In all cases, the Gaussian kernel $K(u)=(2\pi)^{-1/2} \exp(-u^2/2)$ was used.  This kernel satisfies Conditions \ref{cond:C1}--\ref{cond:C5}.  Bandwidths were found for the
standard and proposed approaches through biased cross-validation ({\tt BCV}, {\tt mBCV}), the Sheather-Jones solve-the-equation method ({\tt SJse}, {\tt mSJse}), and the Sheather-Jones method of minimizing the objective function ({\tt SJmin}, {\tt mSJmin}).  As an aspirational target, we also compute the bandwidth that minimizes the  integrated squared error, $\ISE = \int \{ \hat{f}_h(x) - f(x) \}^2 \diff x$, which is known in the simulation study, but would not be known in practice.

Table \ref{table:ISE} shows the bandwidth and integrated squared error averaged over $50$ replicates for each of the six simulation settings: independent and dependent samples for the three true distributions. With independent samples, the integrated autocorrelation times are near $1$ and the performance of proposed methods is comparable to that of the standard methods in terms of chosen bandwidth, density estimate, and  average integrated squared error. 

With Markov chain Monte Carlo samples, the bandwidths chosen by standard methods are consistently much smaller than the aspirational target bandwidth, confirming that the standard methods often produce undersmoothed estimates when the data are dependent \citep{Hart:1990}. In contrast, the proposed methods result in values that are much closer to the aspirational target bandwidth. The proposed methods have consistently smaller average integrated squared error than the standard methods. 
Specifically, the percentage decrease of the average integrated squared error by using the proposed methods instead of the standard methods ranges from 4\% to 57\%, with a median percentage decrease of 25\%. 
After removing the average integrated squared error of the aspirational target, the excess average integrated squared errors of the standard methods are 1$\cdot$4  times to 14$\cdot$4 times greater than those of the corresponding proposed methods.

Many practitioners thin the Markov chain by subsampling to reduce autocorrelation.  Theoretical results show that this thinning hurts the performance of estimators \citep{Geyer:1992,MacEachern:1994}.  The last column of Table \ref{table:ISE} includes results
for thinned samples which retain every 5th observation from the Markov chain.  The density estimate
is a Sheather-Jones solve-the-equation estimate.  The thinned sample method leads to a smaller average integrated squared error than the standard methods when the data are dependent.  However, when the samples are independent, the thinned sample method performs poorly since four-fifths of the data have been discarded.  The proposed methods outperform the thinned sample method in all cases. 

\begin{table}[htbp]
\label{table:ISE}
\caption{
Bandwidth $h$ and integrated squared error ({\ISE}) averaged over 50 replicates, with standard errors in parentheses. Six simulated settings consist of samples from a normal distribution, a two-component mixture normal distribution, and a log-normal distribution, each drawn from an independent sampler and by Markov chain Monte Carlo. The three proposed methods ({\tt mBCV}, {\tt mSJse}, {\tt mSJmin}) are compared to the standard methods ({\tt BCV}, {\tt SJse}, {\tt SJmin}), the aspirational target ({\tt Target}), and a thinned sample method ({\tt Thin}).
}

{%
\footnotesize
\begin{tabular}{cc c ccc ccc c}
\\
& & & \multicolumn{3}{c}{Standard methods} & \multicolumn{3}{c}{Proposed methods} & 
 \\
       & & \multicolumn{1}{c}{\tt Target}    
        & \multicolumn{1}{c}{\tt BCV} & \multicolumn{1}{c}{\tt SJse} & \multicolumn{1}{c}{\tt SJmin} & \multicolumn{1}{c}{\tt mBCV} & \multicolumn{1}{c}{\tt mSJse} & \multicolumn{1}{c}{\tt mSJmin} & \multicolumn{1}{c}{\tt Thin} \\
      \multirow{4}{*}{\parbox{7em}{\centering Normal: \qquad Independent}} 	& \multirow{2}{*}{$h$}	& 0$\cdot$338 & 0$\cdot$343 & 0$\cdot$336 & 0$\cdot$348 & 0$\cdot$344 & 0$\cdot$337 & 0$\cdot$349 & 0$\cdot$459 \\
      	&  	& (0$\cdot$009) & (0$\cdot$002) & (0$\cdot$001) & (0$\cdot$001) & (0$\cdot$002) & (0$\cdot$001) & (0$\cdot$001) & (0$\cdot$003) \\ 
	\vspace{-0.9em} \\
     			& \multirow{2}{*}{\ISE}	& 0$\cdot$087 & 0$\cdot$095 & 0$\cdot$095 & 0$\cdot$095 & 0$\cdot$095 & 0$\cdot$095 & 0$\cdot$095 & 0$\cdot$123 \\ 
      				& 	& (0$\cdot$006) & (0$\cdot$006) & (0$\cdot$006) & (0$\cdot$006) & (0$\cdot$006) & (0$\cdot$006) & (0$\cdot$006) & (0$\cdot$008) \\ 
      \vspace{-0.4em} \\
   \multirow{4}{*}{\parbox{7em}{\centering Normal: \qquad MCMC sample}}   		& \multirow{2}{*}{$h$} 		& 0$\cdot$529 & 0$\cdot$398 & 0$\cdot$228 & 0$\cdot$327 & 0$\cdot$550 & 0$\cdot$475 & 0$\cdot$529 & 0$\cdot$424 \\ 
     			& 	& (0$\cdot$012) & (0$\cdot$009) & (0$\cdot$006) & (0$\cdot$004) & (0$\cdot$005) & (0$\cdot$005) & (0$\cdot$003) & (0$\cdot$005) \\ 
      \vspace{-0.9em} \\
     				& \multirow{2}{*}{\ISE}		& 0$\cdot$347 & 0$\cdot$493 & 0$\cdot$960 & 0$\cdot$589 & 0$\cdot$395 & 0$\cdot$417 & 0$\cdot$391 & 0$\cdot$458 \\ 
      				& 	& (0$\cdot$030) & (0$\cdot$038) & (0$\cdot$077) & (0$\cdot$044) & (0$\cdot$032) & (0$\cdot$032) & (0$\cdot$030) & (0$\cdot$035) \\ 
      \vspace{-0.4em} \\   
   \multirow{4}{*}{\parbox{7em}{\centering Normal mixture: \qquad Independent}}    		& \multirow{2}{*}{$h$} 		& 0$\cdot$183 & 0$\cdot$189 & 0$\cdot$189 & 0$\cdot$195 & 0$\cdot$189 & 0$\cdot$189 & 0$\cdot$196 & 0$\cdot$267 \\ 
      	& 	& (0$\cdot$004) & (0$\cdot$001) & (0$\cdot$000) & (0$\cdot$000) & (0$\cdot$001) & (0$\cdot$000) & (0$\cdot$000) & (0$\cdot$001) \\ 
      \vspace{-0.9em} \\
     				& \multirow{2}{*}{\ISE}		& 0$\cdot$179 & 0$\cdot$188 & 0$\cdot$188 & 0$\cdot$189 & 0$\cdot$188 & 0$\cdot$188 & 0$\cdot$189 & 0$\cdot$266 \\ 
      				& 	& (0$\cdot$009) & (0$\cdot$009) & (0$\cdot$009) & (0$\cdot$009) & (0$\cdot$009) & (0$\cdot$009) & (0$\cdot$010) & (0$\cdot$015) \\ 
      \vspace{-0.4em} \\
   \multirow{4}{*}{\parbox{7em}{\centering Normal mixture: \qquad MCMC sample}}   		& \multirow{2}{*}{$h$}		& 0$\cdot$293 & 0$\cdot$228 & 0$\cdot$159 & 0$\cdot$191 & 0$\cdot$319 & 0$\cdot$301 & 0$\cdot$323 & 0$\cdot$257 \\ 
    			& 	& (0$\cdot$007) & (0$\cdot$004) & (0$\cdot$002) & (0$\cdot$001) & (0$\cdot$002) & (0$\cdot$002) & (0$\cdot$002) & (0$\cdot$002) \\ 
      \vspace{-0.9em} \\
     				& \multirow{2}{*}{\ISE}		& 0$\cdot$900 & 1$\cdot$117 & 1$\cdot$510 & 1$\cdot$246 & 0$\cdot$996 & 0$\cdot$981 & 0$\cdot$991 & 1$\cdot$014 \\ 
      				& 	& (0$\cdot$068) & (0$\cdot$076) & (0$\cdot$089) & (0$\cdot$074) & (0$\cdot$072) & (0$\cdot$070) & (0$\cdot$073) & (0$\cdot$067) \\ 
      \vspace{-0.4em} \\
	\multirow{4}{*}{\parbox{7em}{\centering Log-normal: \qquad Independent}}    		& \multirow{2}{*}{$h$}	& 0$\cdot$125 & 0$\cdot$123 & 0$\cdot$120 & 0$\cdot$125 & 0$\cdot$123 & 0$\cdot$120 & 0$\cdot$125 & 0$\cdot$165 \\ 
      	& 	& (0$\cdot$003) & (0$\cdot$001) & (0$\cdot$000) & (0$\cdot$000) & (0$\cdot$001) & (0$\cdot$000) & (0$\cdot$000) & (0$\cdot$001) \\ 
      \vspace{-0.9em} \\
     				& \multirow{2}{*}{\ISE}		& 0$\cdot$245 & 0$\cdot$260 & 0$\cdot$260 & 0$\cdot$260 & 0$\cdot$260 & 0$\cdot$260 & 0$\cdot$260 & 0$\cdot$339 \\ 
      				& 	& (0$\cdot$015) & (0$\cdot$016) & (0$\cdot$016) & (0$\cdot$016) & (0$\cdot$016) & (0$\cdot$016) & (0$\cdot$016) & (0$\cdot$020) \\ 
      \vspace{-0.4em} \\
   \multirow{4}{*}{\parbox{7em}{\centering Log-normal: \qquad MCMC sample}}    		& \multirow{2}{*}{$h$}	& 0$\cdot$189 & 0$\cdot$155 & 0$\cdot$083 & 0$\cdot$122 & 0$\cdot$213 & 0$\cdot$190 & 0$\cdot$208 & 0$\cdot$156 \\ 
    			& 	& (0$\cdot$004) & (0$\cdot$003) & (0$\cdot$002) & (0$\cdot$001) & (0$\cdot$002) & (0$\cdot$001) & (0$\cdot$001) & (0$\cdot$002) \\ 
      \vspace{-0.9em} \\
     				& \multirow{2}{*}{\ISE}		& 1$\cdot$202 & 1$\cdot$413 & 2$\cdot$988 & 1$\cdot$776 & 1$\cdot$354 & 1$\cdot$326 & 1$\cdot$337 & 1$\cdot$456 \\ 
      				& 	& (0$\cdot$087) & (0$\cdot$090) & (0$\cdot$228) & (0$\cdot$124) & (0$\cdot$098) & (0$\cdot$086) & (0$\cdot$092) & (0$\cdot$097)  
\end{tabular}}

\end{table}

\section{Discussion} \label{sec:discussion}


The proposed approaches also apply to time series that satisfy certain mixing conditions.  Specifically, Theorem \ref{eq:MISE_var_dep} and Corollary \ref{cor:MISE_mAMISE:dep} apply to density estimation for samples from an $\alpha$-mixing time series  \citep{Robinson:1983,Roussas:1988,Liebscher:1996}.  For a $\rho$-mixing time series \citep{Bradley:1983}, Theorem \ref{thm:IACT_geo}, Corollary \ref{cor:geo}, and Theorem \ref{thm:conv_zeta_f} hold, so our proposed bandwidth selection methods can be directly applied.

\pagebreak

%
%
%


\appendix

\section*{Appendix 1}

This section summarizes theorems that connect Markov chain Monte Carlo to the mixing behavior of processes.
The mixing conditions are used in the proofs in Appendix 2.


\begin{theorem} \label{thmA:Jones_2_1} \citep[Thm 2.1]{Jones:2004} Let $\mathcal{F}_i^j$ denote the $\sigma$-field generated by $\{Y_i,\ldots,Y_j\}$ for $i \le j$. If $\{Y_1,\ldots,Y_n\}$ is a Harris ergodic Markov chain with stationary distribution $f$, then the chain is $\alpha$-mixing, i.e., as $n \rightarrow \infty$, 
\begin{equation*} \label{eq:strong_mix}
	\alpha_n \equiv \sup \left\{ |\pr(U \cap V) - \pr(U)\pr(V) |: U \in \mathcal{F}_1^k, \ V \in \mathcal{F}_{k+n}^{\infty}, \ k \ge 1 \right\} \ \rightarrow 0.
\end{equation*}
\end{theorem}

\begin{theorem} \label{thmA:Jones_2_2} \citep[Thm 2.2]{Jones:2004} If $\{Y_1,\ldots,Y_n\}$ is a geometrically ergodic chain and the kernel $P$ satisfies detailed balance with respect to $f$, i.e., 
$f(\diff y) P(y, \diff y') = f( \diff y') P(y', \diff y)$ for $y, y' \in \mathcal{Y}$, then the chain is $\rho$-mixing, i.e., for some $\theta > 0$
\begin{equation} \label{eq:rho_mix}
\rho(n) \equiv \sup \left\{ \corr (U,V): U \in \mathcal{L}^2\left(\mathcal{F}_1^k\right), \ V \in \mathcal{L}^2\left(\mathcal{F}_{k+n}^{\infty}\right), \ k \ge 1 \right\} = O \left( e^{-\theta n} \right)
\end{equation}
as $n \rightarrow \infty$, where $\mathcal{L}^2(\mathcal{F}) = \{W \in \mathcal{F}; E (W^2) < \infty \}$.
\end{theorem}

\begin{lemma} \label{lemmaA:Tran_L1} \cite[Corollary 2.1]{Tran:1989}
Suppose that $\{Y_1,\ldots,Y_n\}$ is $\alpha$-mixing with $\alpha_n=O \left( e^{-sn} \right)$ as $n \rightarrow \infty$ for some $s > 0$. Then, $ \int \left| \hat{f}_h(x) - f_h(x) \right| \diff x \rightarrow 0$ almost surely if $nh (\log n)^{-3} \rightarrow \infty$.
\end{lemma}

\begin{definition} \label{defA:Tran2_1} 
 \cite[Lemma 2.1]{Tran:1990} Let $g(\cdot,\cdot)$ be a nonnegative function on $N \times N$. Specifically, let $g(mp,p) = C^* \ (mp+p)^\theta$ for some $C^* > 0$ and some $\theta \ge 0$. Let $\psi$ be a decreasing function such that $\psi(p) \downarrow 0$ as $p \rightarrow \infty$. Then, the process $\{Y_t\}$ is said to satisfy the strong mixing property in the locally transitive sense with regard to $g$ if
$$
	\gamma(m,p) \equiv \sup  \left\{ |\pr(A \cap B) - \pr(A)\pr(B) |: A \in \mathcal{F}_1^{mp}, \ B \in \mathcal{F}_{(m+1)p+1}^{(m+2)p} \right\}  \le  C \ g(mp,p) \ \psi(p)
$$
for all positive integers $m$ and $p$ and for some constant $C>0$.
\end{definition}

\begin{lemma} \label{lemmaA:TranRemark2_1} \cite[Lemma 2.2, Remark 2.1]{Tran:1990} Assume $\sum_{k=1}^\infty \psi(k)^{\delta/(2+\delta)} < \infty$ for some $\delta > 0$. If the kernel $K$ satisfies Conditions \ref{cond:C1}--\ref{cond:C5} and $\{Y_1,\ldots,Y_n\}$ satisfy the strong mixing property in the locally transitive sense, then $ \var\{\hat{f}_h(x)\} = O \left( n^{-1} h^{-1 - \delta/(2+\delta)} \right)$ as $n \rightarrow \infty$.
\end{lemma}

\begin{lemma} \label{lemmaA:R_r} For any $h > 0$, $\epsilon >0$, and integer $r \ge 1$, 
\begin{align*}
\phantom{BBB} & \int |t|^r K(t) \int_0^1 \left| f^{(r)}(x-htw) - f^{(r)}(x) \right| (1-w)^{r-1} \diff w  \diff t  \\
& \le \ \max_{|y| < \epsilon } | f^{(r)}(x-y) - f^{(r)}(x) | \int |t|^r K(t) \diff t 
+ \max_{|t| \ge \frac{\epsilon}{h}} \{ |t|^{r+1} K(t) \} \int | f^{(r)}(y) | \diff y \frac{1}{\epsilon} \hspace{50em} \\
& \qquad   + \int I \left(|t| \ge \frac{\epsilon}{h} \right) |t|^r K(t) \ \diff t | f^{(r)}(x) |. 
\end{align*}
\end{lemma}

\begin{lemma} \label{lemmaA:r_x_h} Suppose the kernel $K$ satisfies Conditions \ref{cond:C1}--\ref{cond:C5} and define 
$$
	r(x,h) = \int t K^2(t) \int_0^1 \{ f^{'}(x-htw) - f^{'}(x) \} \diff w \diff t.
$$
Then, (a) $|r(x,h)|<\infty$ almost everywhere for any $h > 0$, (b) $r(x,h)$ converges to zero almost everywhere as $h \rightarrow 0$, and (c) $\lim_{h \rightarrow 0} \int r(x,h) \diff x = 0$.
\end{lemma}

\noindent {\it Proof.}
\begin{align}
\phantom{000} |r(x,h)| &  
\le \int |t| K^2(t) \int_0^1 \left| f^{'}(x-htw) - f^{'}(x) \right| \diff w \diff t 
\nonumber \hspace{50em} \\ 
 & 
\le \max_{t^*} K(t^*) \int |t| K(t) \int_0^1 \left| f^{'}(x-htw) - f^{'}(x) \right| \diff w \diff t  \nonumber \\
& \le \max_{t^*} K(t^*) \left[ \max_{|y| < \epsilon } | f^{'}(x-y) - f^{'}(x) | \int |t| K(t) \diff t 
+ \max_{|t| \ge \frac{\epsilon}{h}} \{ |t|^{2} K(t) \} \int | f^{'}(y) | \diff y \frac{1}{\epsilon} 
\right. \nonumber \\
&  \left. 
\qquad + \int I \left(|t| \ge \frac{\epsilon}{h} \right) |t| K(t) \ \diff t | f^{'}(x) | \right]. \label{eq:lemma:A3}
\end{align}
The last inequality holds with Lemma \ref{lemmaA:R_r} with $r=1$. In \eqref{eq:lemma:A3}, the first term equals zero with Conditions \ref{cond:C2} and \ref{cond:C4} for any $\epsilon > 0$; the second term is finite for any $h$ and goes to zero as $h \rightarrow 0$ with Conditions \ref{cond:C1} and \ref{cond:C3}; the third term is finite for any $h$ and goes to zero as $h \rightarrow 0$ with Condition \ref{cond:C2} and because Condition \ref{cond:C3} implies $|f'(x)| < \infty$ almost everywhere. Therefore, $|r(x,h)| < \infty$ for any $h > 0$ and $r(x,h) \rightarrow 0$ almost everywhere as $h \rightarrow 0$. Then, by the bounded convergence theorem $\lim_{h \rightarrow 0} \int r(x,h) \diff x = 0$.

\begin{lemma} \label{lemmaA:qr_x_h} Suppose the kernel $K$ satisfies Conditions \ref{cond:C1}--\ref{cond:C5} and define 
$$
	q_r(x,h) = \int t^r K(t) \int_0^1 \left\{ f^{(r)}(x-htw) - f^{(r)}(x) \right\} (1-w)^{r-1} \diff w \diff t
$$
for an integer $r \ge 1$. Then, (a) $|q_r(x,h)|<\infty$ almost everywhere for any $h > 0$, (b) $q_r(x,h)$ converges to zero almost everywhere as $h \rightarrow 0$, and (c) $\lim_{h \rightarrow 0} \int q_r(x,h) \diff x = 0$.
\end{lemma}

\noindent {\it Proof.}
From Lemma \ref{lemmaA:R_r},
\begin{align*}
|q_r(x,h)| \le &  
\max_{|y| < \epsilon } | f^{(r)}(x-y) - f^{(r)}(x) | \int |t|^r K(t) \diff t + \max_{|t| \ge \frac{\epsilon}{h}} \{ |t|^{r+1} K(t) \} \int | f^{(r)}(y) | \diff y \frac{1}{\epsilon} \\
& \ \ + \int I \left(|t| \ge \frac{\epsilon}{h} \right) |t|^r K(t) \ \diff t | f^{(r)}(x) |.
\end{align*}
The first term equals zero with Conditions \ref{cond:C2} and \ref{cond:C4} for any $\epsilon > 0$; the second term is finite for any $h$ and goes to zero as $h \rightarrow 0$ with Conditions \ref{cond:C1} and \ref{cond:C3}; the third term is finite for any $h$ and goes to zero as $h \rightarrow 0$ with Condition \ref{cond:C2} and because Condition \ref{cond:C3} implies $|f^{(r)}(x)| < \infty$ almost everywhere. Therefore, $|q_r(x,h)| < \infty$ for any $h > 0$ and $q_r(x,h) \rightarrow 0$ almost everywhere as $h \rightarrow 0$. Then, by the bounded convergence theorem $\lim_{h \rightarrow 0} \int q_r(x,h) \diff x = 0$.


\section*{Appendix 2}

This section provides the proofs of theorems in the main text. \\

\noindent {\it Proof of Theorem \ref{thm:tau_K}} \\ 

Using Taylor's theorem with the integral form of the remainder,
$$ E \{ K^2_h(x-Y_1) \} = \frac{1}{h} \R(K) f(x) - \int t K^2(t) \diff t \ f^{'}(x)
- r(x,h) = h^{-1} \R(K) f(x)  + O(1) $$
as $h \rightarrow 0$
because $\int t K^2(t) \diff t \le \max_{t} K(t) \int t K(t) \diff t < \infty$ and $|r(x,h)| < \infty$ from Lemma \ref{lemmaA:r_x_h}.
Here, $\mu_1=0$ for $K$ satisfying Condition \ref{cond:C5}. By using Taylor's theorem with the integral form of the remainder again, 
$$ E \left\{ K_h(x-Y_1) \right\}^2 = f^2(x) - 2h f(x) q_1(x,h) + h^2 q_1(x,h)^2 = f^2(x) + O(h) $$
as $h \rightarrow 0$ because $ q_1(x,h) \equiv \int t K(t) \int_0^1 \left\{ f^{'}(x-htw) - f^{'}(x) \right\} \diff w < \infty$ from Lemma \ref{lemmaA:qr_x_h}. Therefore, 
$ n^{-1} \var\{K_h(x-Y_1)\} = (nh)^{-1} \R(K) f(x) + o \left(n^{-1}h^{-1} \right)$ as $n \rightarrow \infty$.

By Theorem \ref{thmA:Jones_2_1}, $\{Y_n\}$ is $\alpha$-mixing, which implies strong mixing in the locally transitive sense in Definition \ref{defA:Tran2_1}. Then, by Lemma \ref{lemmaA:TranRemark2_1}, $\var\{\hat{f}_h(x)\} = O \left( n^{-1} h^{-1 - \delta/(2+\delta)} \right)$ as $n \rightarrow \infty$. 
From \eqref{eq:newvar}, $\var\{\hat{f}_h(x)\} = n^{-1} \var\{K_h(x-Y_1)\}\tau_n(K_{h,x})$. 
Therefore, $ \tau_n(K_{h,x})=O \left( h^{-\delta/(2+\delta)} \right)$ as $n \rightarrow \infty.$ \hfill $\square$ \\

\noindent {\it Proof of Theorem \ref{thm:IACT_geo}} \\ 

It is straightforward to show that $h^{\frac{1}{2}} K_h(x-Y_1)$ is measurable with respect to $\mathcal{F}_1^1$, $h^{\frac{1}{2}} K_h(x-Y_{t+1})$ is \phantom{0}
measurable with respect to $\mathcal{F}_{t+1}^{\infty}$, and $h^{\frac{1}{2}} K_h(x-Y_1)$ is square-integrable, i.e., 
$$ E \left[ \left\{ h^{\frac{1}{2}} K_h(x-Y_1) \right\}^2 \right] = h E \{ K^2_h(x-Y_1) \} = \R(K) f(x) - h \int t K^2(t) \diff t f^{'}(x) - h r(x,h) < \infty. $$
Therefore, by Theorem \ref{thmA:Jones_2_2} there exists a $\theta > 0$ such that $ \left| \corr \left\{h^{\frac{1}{2}} K_h(x-Y_1), h^{\frac{1}{2}} K_h(x-Y_{t+1}) \right\} \right| \le \rho(t)$ defined in \eqref{eq:rho_mix}.
Then, 
\begin{align*}
\phantom{000000}  |\tau_n(K_{h,x})|   
& \le \sum_{t=-(n-1)}^{n-1} | \corr \left\{K_h(x-Y_1),K_h(x-Y_{t+1})\right\} | \hspace{50em} \\
& = 1 + 2 \sum_{t=1}^{n-1} | \corr \left\{ K_h(x-Y_1),K_h(x-Y_{t+1}) \right\} | \le 1 + 2 \sum_{t=1}^{n-1} \rho(t).
\end{align*}
The equality holds because $\{Y_1,\ldots,Y_n\}$ is stationary. From Theorem \ref{thmA:Jones_2_2}, $\rho(t)=C \exp(-\theta t)$ for a constant $C$ and any $t > 0$. Therefore, the sum of geometric series $\sum_{t=1}^{n-1} \rho(t)$ converges because $\exp(-\theta t) < 1$ for any $\theta, t > 0$. \hfill $\square$ \\

\noindent {\it Proof of Theorem \ref{thm:MISE_MCMC}} \\ 

Using Taylor's theorem with the integral form of the remainder,
\begin{align*}
& n^{-1} \int E \left\{ K_h^2(x-Y_1) \right\} \ \tau_n(K_{h,x}) \diff x \hspace{50em} \\
& = (nh)^{-1} \R(K) \int \tau_n(K_{h,x}) f(x) \diff x - n^{-1} \int t \ K^2(t) \diff t \int f^{'}(x) \ \tau_n(K_{h,x}) \diff x - n^{-1} \int r(x,h) \ \tau_n(K_{h,x}) \diff x \\
& = (nh)^{-1} \R(K) \zeta_f(K_h) - n^{-1} h^{-\frac{\delta}{2+\delta}} \int t \ K^2(t) \diff t \int f^{'}(x) \ h^{\frac{\delta}{2+\delta}} \tau_n(K_{h,x}) \diff x \\
& \ \ - n^{-1} h^{-\frac{\delta}{2+\delta}} \int r(x,h) \ h^{\frac{\delta}{2+\delta}} \tau_n(K_{h,x}) \diff x  \\
& = (nh)^{-1} \R(K) \zeta_f(K_h) + O \left(n^{-1} h^{-\frac{\delta}{2+\delta}} \right)
\end{align*}
as $n \rightarrow \infty$ because $h^{\delta/(2+\delta)} \tau_n(K_{h,x}) = O(1)$ from Theorem \ref{thm:tau_K} and $\int r(x,h) \ h^{\delta/(2+\delta)} \tau_n(K_{h,x}) \diff x \rightarrow 0$ as $h \rightarrow 0$ by the bounded convergence theorem with Lemma \ref{lemmaA:r_x_h}.

Using Taylor's theorem again,
\begin{align*}
\phantom{000000} & n^{-1} \int \left[ E\{K_h(x-Y_1)\} \right]^2 \tau_n(K_{h,x}) \diff x \hspace{50em} \\
& = n^{-1} \int \left[ f(x) - h \int t K(t) \int_0^1 \left\{ f^{'}(x-htw) - f^{'}(x) \right\} \diff w \diff t \right]^2 \tau_n(K_{h,x}) \diff x 
& = n^{-1} h^{-\frac{\delta}{2+\delta}} \int f^2(x) \ h^{\frac{\delta}{2+\delta}} \tau_n(K_{h,x}) \diff x  - 2 n^{-1} h^{1-\frac{\delta}{2+\delta}} \int f(x) q_1(x,h) \ h^{\frac{\delta}{2+\delta}} \tau_n(K_{h,x}) \diff x \hspace{50em} \\
& \ \ + n^{-1} \ h^{2-\frac{\delta}{2+\delta}} \int \{ q_1(x,h) \}^2 h^{\frac{\delta}{2+\delta}} \tau_n(K_{h,x}) \diff x \\
& = n^{-1} h^{-\frac{\delta}{2+\delta}} \int f^2(x) \ h^{\frac{\delta}{2+\delta}} \tau_n(K_{h,x}) \diff x + o \left( n^{-1} \right)
\end{align*}
as $n \rightarrow \infty$ because $h^{\delta/(2+\delta)} \tau_n(K_{h,x}) = O(1)$ from Theorem \ref{thm:tau_K} and $\int f(x) q_1(x,h) h^{\frac{\delta}{2+\delta}} \tau_n(K_{h,x}) \diff x \rightarrow 0$ as $h \rightarrow 0$ by the bounded convergence theorem with Lemma \ref{lemmaA:r_x_h} when $|\R(f)| < \infty$.
Therefore,
\begin{align*}
\phantom{00000}  \int \var \{\hat{f}_h(x)\} \diff x &  = \frac{1}{n} \int \var \left\{ K_h(x-Y_1) \right\} \tau_n(K_{h,x}) \diff x  \hspace{50em} \\
& = \frac{1}{n} \int E \left\{ K_h^2(x-Y_1) \right\} \tau_n(K_{h,x}) \diff x - \frac{1}{n} \int \left[ E\{K_h(x-Y_1)\} \right]^2 \tau_n(K_{h,x}) \diff x \\
&  = \frac{1}{nh} \R(K) \zeta_f(K_h) + O \left( n^{-1} h^{-\frac{\delta}{2+\delta}} \right) \text{ as } n \rightarrow \infty. \hspace{11em} \square
\end{align*}


\noindent {\it Proof of Theorem \ref{thm:conv_zeta_f}}  

\begin{align*}
& \int \tau_n(K_{h,x}) f_h(x) - \int \tau_n(K_{h,x}) \hat{f}_h(x) \diff x \le \int \left| \tau_n(K_{h,x}) \right| \ \left| f_h(x) - \hat{f}_h(x) \right| \diff x \\
& \le \int  \left( 1+2 \ C \ \frac{e^{-\theta}-e^{-\theta n}}{1-e^{-\theta}} \right) \ \left| f_h(x) - \hat{f}_h(x) \right| \diff x \ \text{ for some constants } \theta >0 \text{ and } C > 0 \\
& \le \int  \left( 1+2 \ C \ \frac{e^{-\theta}}{1-e^{-\theta}} \right) \ \left| f_h(x) - \hat{f}_h(x) \right| \diff x 
= \left( 1+2 \ C \ \frac{e^{-\theta}}{1-e^{-\theta}} \right) \ \int   \left| f_h(x) - \hat{f}_h(x) \right| \diff x.
\end{align*}
The second inequality holds by Theorem \ref{thm:IACT_geo}. By Lemma \ref{lemmaA:Tran_L1}, it goes to zero almost surely.  \hfill $\square$

\bibliographystyle{biometrika}
\bibliography{MasterVer}

\end{document}